\PassOptionsToPackage{unicode}{hyperref}
\PassOptionsToPackage{hyphens}{url}
\documentclass[a4paper,oneside]{article}

\usepackage[
  top=2.5cm,
  bottom=2.5cm,
  left=2.5cm,
  right=2.5cm
]{geometry}
\usepackage{xcolor}
\usepackage{amsmath,amssymb}
\usepackage{iftex}
\ifPDFTeX
  \usepackage[T1]{fontenc}
  \usepackage[utf8]{inputenc}
  \usepackage{textcomp} 
\else 
  \usepackage{unicode-math} 
  \defaultfontfeatures{Scale=MatchLowercase}
  \defaultfontfeatures[\rmfamily]{Ligatures=TeX,Scale=1}
\fi
\usepackage{lmodern}
\ifPDFTeX\else
\fi
\IfFileExists{upquote.sty}{\usepackage{upquote}}{}
\IfFileExists{microtype.sty}{
  \usepackage[]{microtype}
  \UseMicrotypeSet[protrusion]{basicmath} 
}{}
\makeatletter
\@ifundefined{KOMAClassName}{
  \IfFileExists{parskip.sty}{%
    \usepackage{parskip}
  }{
    \setlength{\parindent}{0pt}
    \setlength{\parskip}{6pt plus 2pt minus 1pt}}
}{
  \KOMAoptions{parskip=half}}
\makeatother
\usepackage{longtable,booktabs,array}
\usepackage{calc} 
\usepackage{etoolbox}
\makeatletter
\patchcmd\longtable{\par}{\if@noskipsec\mbox{}\fi\par}{}{}
\makeatother
\IfFileExists{footnotehyper.sty}{\usepackage{footnotehyper}}{\usepackage{footnote}}
\makesavenoteenv{longtable}
\usepackage{graphicx}
\makeatletter
\newsavebox\pandoc@box
\newcommand*\pandocbounded[1]{
  \sbox\pandoc@box{#1}%
  \Gscale@div\@tempa{\textheight}{\dimexpr\ht\pandoc@box+\dp\pandoc@box\relax}%
  \Gscale@div\@tempb{\linewidth}{\wd\pandoc@box}%
  \ifdim\@tempb\p@<\@tempa\p@\let\@tempa\@tempb\fi
  \ifdim\@tempa\p@<\p@\scalebox{\@tempa}{\usebox\pandoc@box}%
  \else\usebox{\pandoc@box}%
  \fi%
}
\def\fps@figure{htbp}
\makeatother
\usepackage{svg}
\ifLuaTeX
  \usepackage{luacolor}
  \usepackage[soul]{lua-ul}
\else
  \usepackage{soul}
\fi
\usepackage{bookmark}
\IfFileExists{xurl.sty}{\usepackage{xurl}}{} 
\hypersetup{
  hidelinks,
  pdfcreator={LaTeX via pandoc}}

\author{}
\date{}

\begin{document}

\begin{center}
{\Large\bfseries
Physics-Informed Latent Neural Operator for Three-Dimensional
Compressor Cascade Flow Prediction\textsuperscript{\dag}
\par}

\vspace{0.8em}

\textsuperscript{1}Yuling Han;
\textsuperscript{1}Zhihui Li\textsuperscript{*};
\textsuperscript{1}Zhibin Yu

\vspace{0.5em}

\emph{\textsuperscript{1}School of Engineering,
University of Liverpool, UK}

\end{center}

\begingroup
\renewcommand{\thefootnote}{\dag}
\footnotetext{This paper was presented at the AI \& Fluids 2026 Conference.}
\endgroup

\textbf{Abstract}

A physics-informed latent neural operator framework is proposed for
three-dimensional compressor cascade flow prediction. The proposed
method combines a Transolver-based latent encoder, which extracts
compact global latent representations of flow conditions from
point-cloud CFD data, with a coordinate-based PINNs decoder to
reconstruct continuous flow fields while incorporating physics-informed
residual constraints during training. The results show that the proposed
framework can accurately reconstruct pressure and velocity distributions
of complex three-dimensional cascade flows and maintains good prediction
capability under previously unseen operating condition. The present
method demonstrates promising potential for efficient CFD surrogate
modelling and aerodynamic prediction in turbomachinery applications.


\textbf{1 Introduction}

High-fidelity computational fluid dynamics (CFD) simulations are widely
used in turbomachinery aerodynamic analysis and compressor cascade
design. However, accurate three-dimensional CFD simulations usually
require large-scale numerical discretisation and iterative solvers,
resulting in substantial computational cost and long simulation times.
These limitations become increasingly significant in applications
involving aerodynamic optimisation and multi-condition flow evaluations.

To reduce the computational expense of conventional CFD simulations,
various deep learning-based surrogate modelling approaches have been
developed in recent years. Early studies mainly employed convolutional
neural networks (CNNs) for flow-field prediction on structured grids
{[}1--3{]}. To improve geometric flexibility, graph neural networks
(GNNs) and neural operator methods were subsequently introduced for PDE
learning on irregular domains {[}4--6{]}. In particular, neural operator
frameworks provide a promising direction for learning mappings between
physical fields under different operating conditions.

More recently, transformer-based architectures have attracted increasing
attention in scientific machine learning and CFD applications due to
their capability for capturing long-range spatial correlations. Among
these approaches, Transolver {[}7{]} introduced a transformer-based PDE
learning framework for general physical domains using slice attention
mechanisms. However, existing Transolver-based approaches remain
primarily data-driven and mainly perform discrete point-wise flow-field
regression without explicitly constructing continuous differentiable
field representations. As a result, incorporating physics-informed PDE
residual constraints into transformer-based CFD surrogate models remains
challenging, and the extrapolation capability of purely data-driven
transformer models is still relatively limited for complex
three-dimensional turbomachinery flows.

Physics-informed neural networks (PINNs) {[}8{]} provide an effective
approach for incorporating governing equations into deep learning
frameworks through PDE residual constraints. However, conventional PINNs
frameworks often suffer from optimization difficulty and slow
convergence when applied to large-scale three-dimensional CFD problems.
In addition, directly solving complex turbulent flow fields using purely
physics-driven PINNs models remains computationally challenging for
practical turbomachinery applications.

To address these challenges, the present study proposes a
physics-informed latent neural operator framework for three-dimensional
compressor cascade flow prediction on CFD point clouds. The proposed
framework combines a Transolver-based latent encoder with a
coordinate-based PINNs decoder to reconstruct continuous flow fields
from latent aerodynamic representations. Physics-informed residual
constraints based on the continuity and momentum equations are further
incorporated during training to improve physical consistency and
extrapolation capability. In addition, prediction task under unseen
attack-angle condition is investigated to evaluate the generalization
performance of the proposed framework.

\textbf{2 Numerical Method}

\textbf{2.1 CFD Simulation and Validation}

The CFD dataset used in the present study was generated from
three-dimensional compressor cascade simulations. The computational
domain consists of a three-dimensional compressor cascade geometry, and
all flow-field data were obtained using ANSYS Fluent. The simulations
were performed on structured meshes in order to accurately capture the
complex geometric boundaries and three-dimensional flow structures
within the cascade passage{[}9{]}.

The computational mesh contains approximately 476,736 cells and 501,168
mesh points. The regular mesh distribution enables refined spatial
resolution near blade surfaces and regions with strong flow gradients,
while maintaining computational efficiency in relatively smooth flow
regions. The resulting CFD solutions were subsequently converted into
regular point-cloud representations for deep learning-based flow-field
prediction.

To verify the reliability of the CFD simulations, the predicted static
pressure coefficient distribution on the blade surface was compared with
available experimental data. The static pressure coefficient is defined
as:

\begin{equation}
C_p = \frac{p-p_{\mathrm{ref}}}
{\frac{1}{2}\rho U_{\mathrm{ref}}^2}
\label{eq:cp}
\end{equation}

where \(p\) denotes the local static pressure, \(p_{ref}\) represents
the reference pressure, \(\rho\) is the fluid density, and \(U_{ref}\)
is the reference inlet velocity.

Figures 1(a) and 1(b) present the comparisons of the static pressure
coefficient distributions at the 5.4\% and 50\% spanwise sections,
respectively. Experimental measurements are represented by open-circle
markers, while CFD results are plotted using solid lines. Results
corresponding to the same attack angle are shown using the same colour.

Overall, the CFD predictions show good agreement with the
experimental data on both the suction side and pressure side of the
blade surface for different attack angles. Compared with the
\(50\%\) spanwise location, relatively larger discrepancies are
observed near the \(5.4\%\) spanwise section due to the more complex
near-wall flow structures close to the endwall region. Nevertheless, the
major pressure variation trends are still accurately captured at both
spanwise locations, indicating that the present CFD setup provides
reliable flow-field data for subsequent deep learning model training.

\begin{center}
\includegraphics[trim=0 9cm 0 0, clip, width=\textwidth]{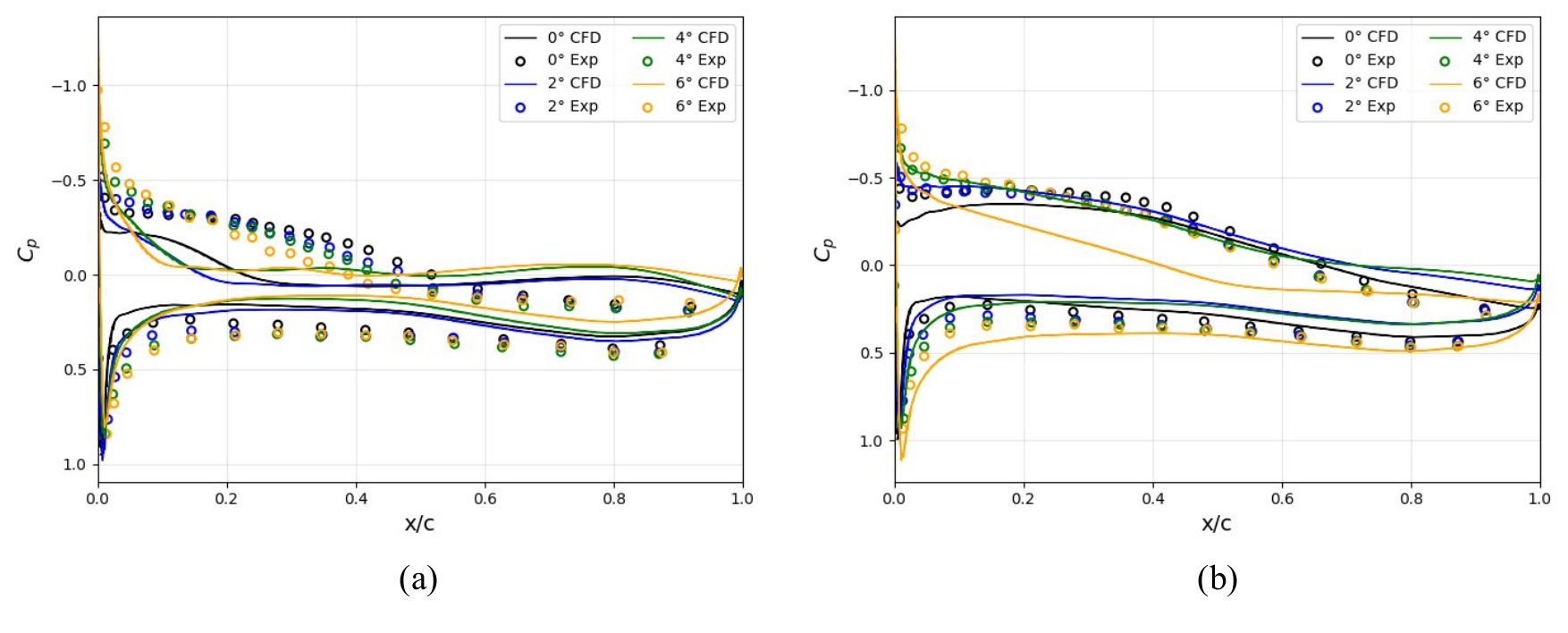}
\end{center}

\emph{Figure 1. Comparison of static pressure coefficient distributions
between CFD predictions and experimental measurements at different
attack angles: (a) 5.4\% spanwise section and (b) 50\% spanwise
section.}

\textbf{2.2 CFD Point-Cloud Representation}

The CFD solutions obtained from \emph{ANSYS Fluent} were converted into
point-cloud representations for deep learning model training.
Specifically, the flow-field data were exported from \emph{ParaView} by
directly saving the spatial point information and corresponding flow
variables from the CFD results.

Each point in the computational domain contains the three-dimensional
spatial coordinates together with the wall-distance information and
operating-condition features. Therefore, the input feature vector of the
\(i\)-th point can be represented as:

\begin{equation}
\mathbf{x}_i =
\left[
x_i,\,
y_i,\,
z_i,\,
d_i,\,
\sin(\alpha),\,
\cos(\alpha)
\right]
\label{eq:input}
\end{equation}

where \(x_{i}\), \(y_{i}\), and \(z_{i}\) denote the spatial coordinates
of the \(i\)-th point, \(d_{i}\) represents the wall distance, and
\(\alpha\) denotes the attack angle. The corresponding target flow
variables are defined as:

\begin{equation}
\mathbf{q}_i =
\left[
u_i,\,
v_i,\,
w_i,\,
p_i,\,
\nu_{\mathrm{eff},i}
\right]
\label{eq:output}
\end{equation}

where \(u_{i}\), \(v_{i}\), and \(w_{i}\) denote the velocity
components, \(p_{i}\) is the static pressure, and \(\nu_{eff,i}\)
represents the effective viscosity. All input and output variables were
normalized using the statistical properties of the training dataset
before model training.

\textbf{2.3 Framework Architecture}

Figure 2 illustrates the overall architecture of the proposed
physics-informed latent neural operator framework. The framework
consists of a Transolver-based latent encoder and a coordinate-based
PINNs decoder for continuous three dimensional flow field
reconstruction.

The Transolver encoder receives geometric point features and
operating-condition information as input and maps the large scale CFD
point-cloud data into a compact latent aerodynamic representation. The
encoder input consists of normalized spatial coordinates, wall-distance
information, and attack angle encoding features. Through multiple
transformer layers with slice attention mechanisms, the encoder captures
long-range spatial correlations and global aerodynamic characteristics
within the compressor cascade flow field.

The latent representation generated by the encoder can be expressed as:

\begin{equation}
\mathbf{z} = \mathcal{E}(\mathbf{x},\mathbf{f})
\label{eq:encoder}
\end{equation}

where \(\mathcal{E}\) denotes the Transolver encoder, \(x\) represents
the geometric point features, \(f\) denotes the attack-angle encoding,
and \(\mathbf{z}\) is the latent aerodynamic representation.

To reconstruct continuous flow fields from the latent representation, a
coordinate-based PINNs decoder was employed. The decoder receives
query-point coordinates together with the latent representation as input
and predicts the corresponding flow variables at arbitrary spatial
locations. The decoder mapping can be written as:

\begin{equation}
\widehat{\mathbf{q}}_i =
\mathcal{D}\left(\mathbf{x}_i,\mathbf{z}\right)
\label{eq:decoder}
\end{equation}

where \(\mathcal{D}\) denotes the coordinate-based decoder and
\(\widehat{\mathbf{q}}\) represents the predicted flow variables.

The decoder predicts the corresponding flow variables at query point
locations. The predicted point-wise variables are then post-processed
for flow-field visualization and comparison with CFD reference results.

\begin{center}
\includegraphics[trim=0 7cm 0 0, clip, width=\textwidth]{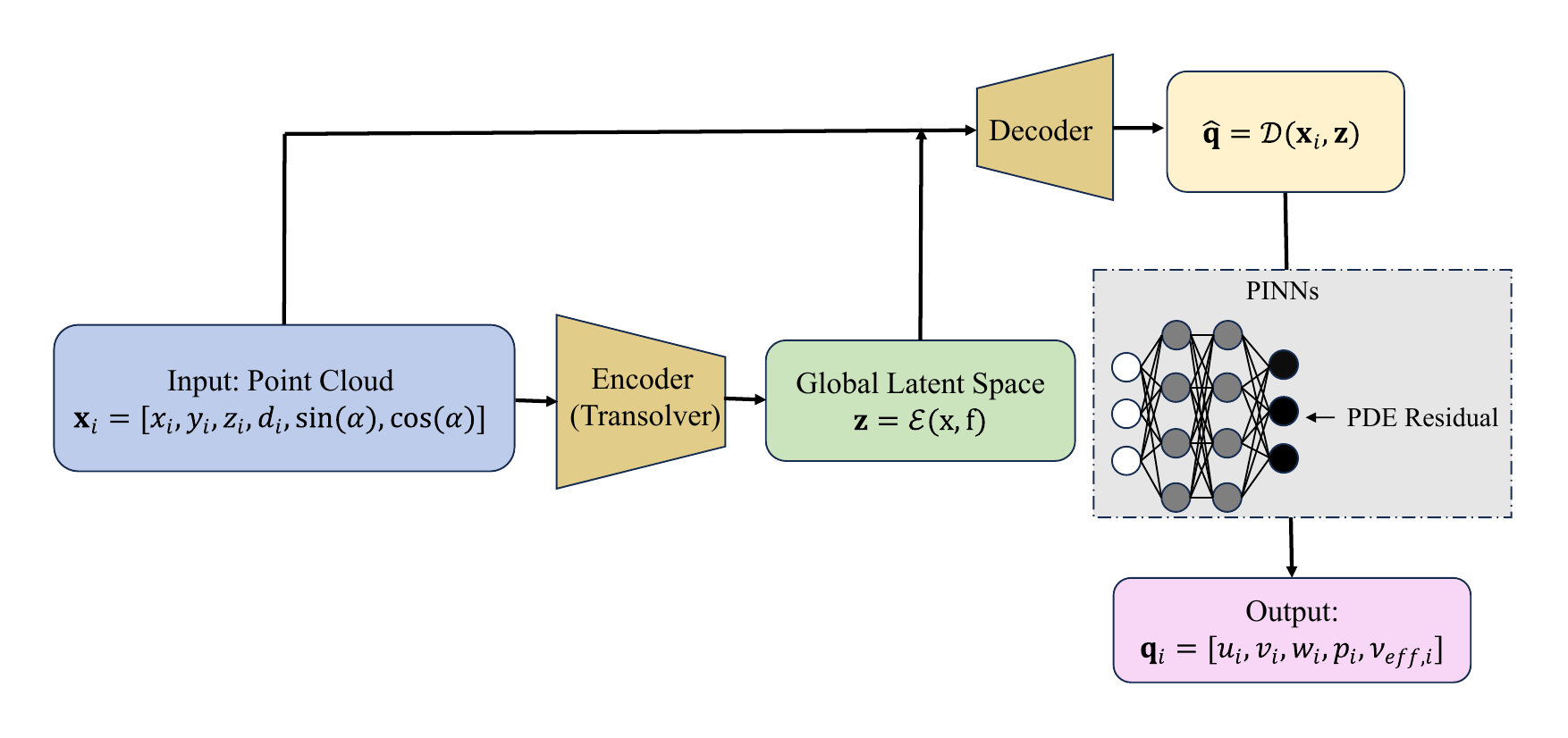}
\end{center}

\begin{center}
\emph{Figure 2. Framework architecture for CFD field prediction}
\end{center}

\textbf{2.4 Physics-Informed Training Strategy}

To improve the physical consistency and extrapolation capability of the
proposed framework, physics-informed residual constraints were
incorporated during training. The continuity equation is defined as:

\begin{equation}
\nabla \cdot \mathbf{u} = 0
\label{eq:continuity}
\end{equation}

The momentum residual equations can be written as:

\begin{equation}
(\mathbf{u}\cdot\nabla)\mathbf{u}
+ \nabla p
- \nabla\cdot\left[
\nu_{\mathrm{eff}}
\left(
\nabla\mathbf{u}
+ (\nabla\mathbf{u})^{T}
\right)
\right]
= 0
\label{eq:momentum}
\end{equation}

During training, automatic differentiation was employed to compute PDE
residuals with respect to the spatial coordinates. The overall loss
function consists of the data reconstruction loss and physics-informed
residual losses:

\begin{equation}
\mathcal{L}
=
\mathcal{L}_{\mathrm{data}}
+
\lambda_{c}\mathcal{L}_{\mathrm{cont}}
+
\lambda_{m}\mathcal{L}_{\mathrm{mom}}
\label{eq:loss}
\end{equation}

where \(\mathcal{L}_{data}\), \(\mathcal{L}_{cont}\), and
\(\mathcal{L}_{mom}\) denote the data loss, continuity residual loss,
and momentum residual loss, respectively. The weighting coefficients
\(\lambda_{c}\) and \(\lambda_{m}\) control the contributions of the
physics-informed constraints during training. To improve computational
efficiency for large-scale CFD point clouds, random point subsampling
was employed during training. The model parameters were optimized using
the \emph{AdamW} optimizer, and all training procedures were implemented
using the \emph{PyTorch} deep learning framework.

\textbf{2.5 Training Datasets and Task Construction}

To investigate the capability of the proposed framework for predicting
compressor cascade flow fields under unknown operating condition, CFD
datasets with multiple attack angles were constructed. The training
dataset consisted of flow-field samples with attack angles of
\(1^{\circ}\), \(2^{\circ}\), \(3^{\circ}\), \(4^{\circ}\),
\(6^{\circ}\), \(7^{\circ}\), \(8^{\circ}\), and \(9^{\circ}\). The flow
field corresponding to the unseen \(5^{\circ}\) attack-angle condition
was employed as the testing dataset.

All CFD datasets were converted into point-cloud representations
containing spatial coordinates, wall-distance information, and
corresponding flow variables. During training and evaluation, the same
data preprocessing and normalization procedures were consistently
applied across all attack-angle conditions.

\textbf{2.6 Computing Facility}

All training and testing procedures in this study were carried out on
the high-performance computing (HPC) platform Barkla at the University
of Liverpool with \emph{Nvidia} H100 GPU card.

\textbf{3 Results}

The capability of the proposed framework was evaluated using the unseen
\(5^{\circ}\) attack-angle condition, which lies within the range of the
training dataset. Fig.3 presents the comparison of the three dimensional
flow-field predictions between the CFD reference solution and the
proposed framework.

\begin{center}
\includegraphics[trim=0 7cm 0 0, clip, width=\textwidth]{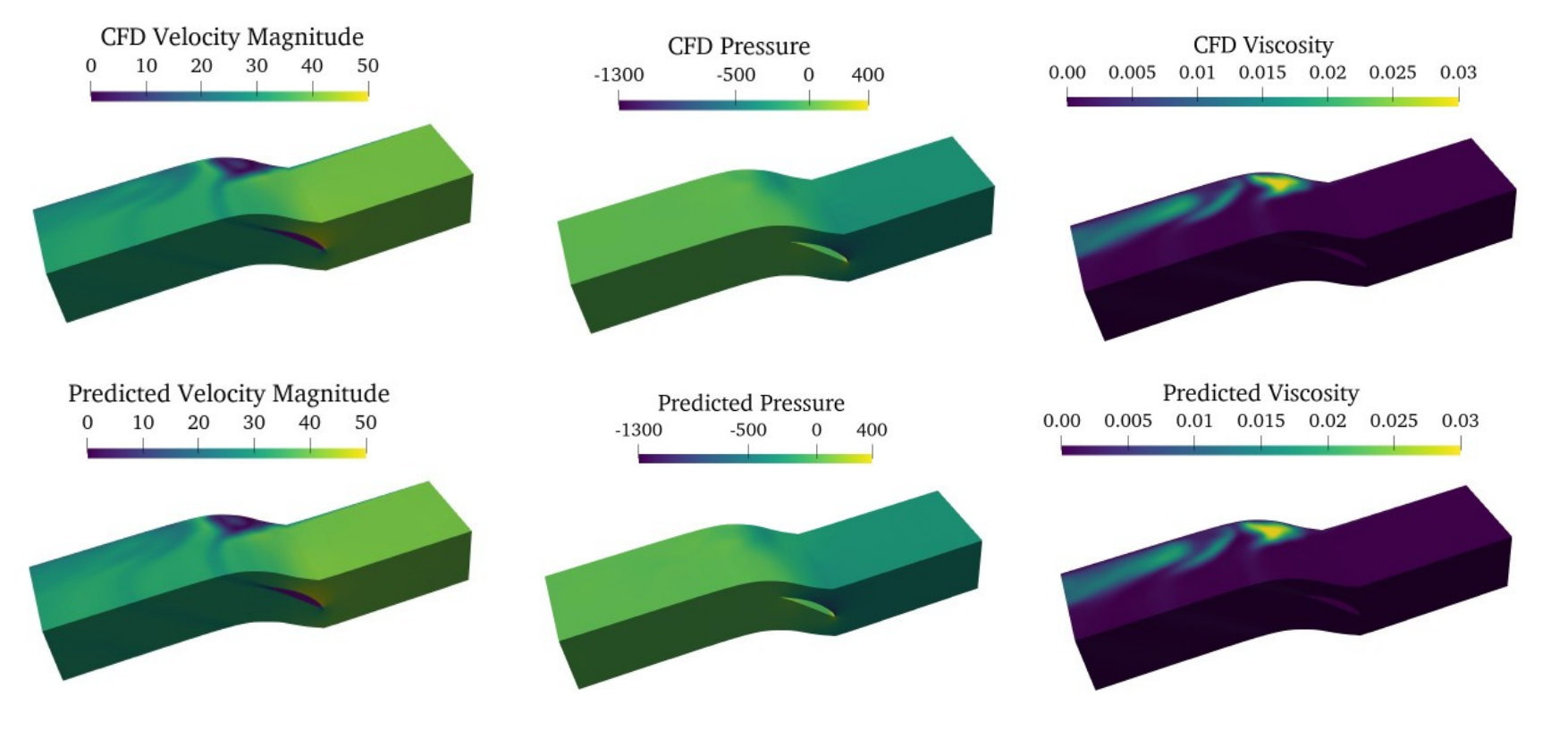}
\end{center}

\emph{Figure 3. Comparison of three-dimensional prediction results at
the} \(5^{\circ}\)\emph{attack-angle condition between the CFD reference
solution and the proposed framework: (a) velocity magnitude, (b)
pressure, and (c) effective viscosity}

\begin{center}
\includegraphics[trim=0 9cm 0 0, clip, width=\textwidth]{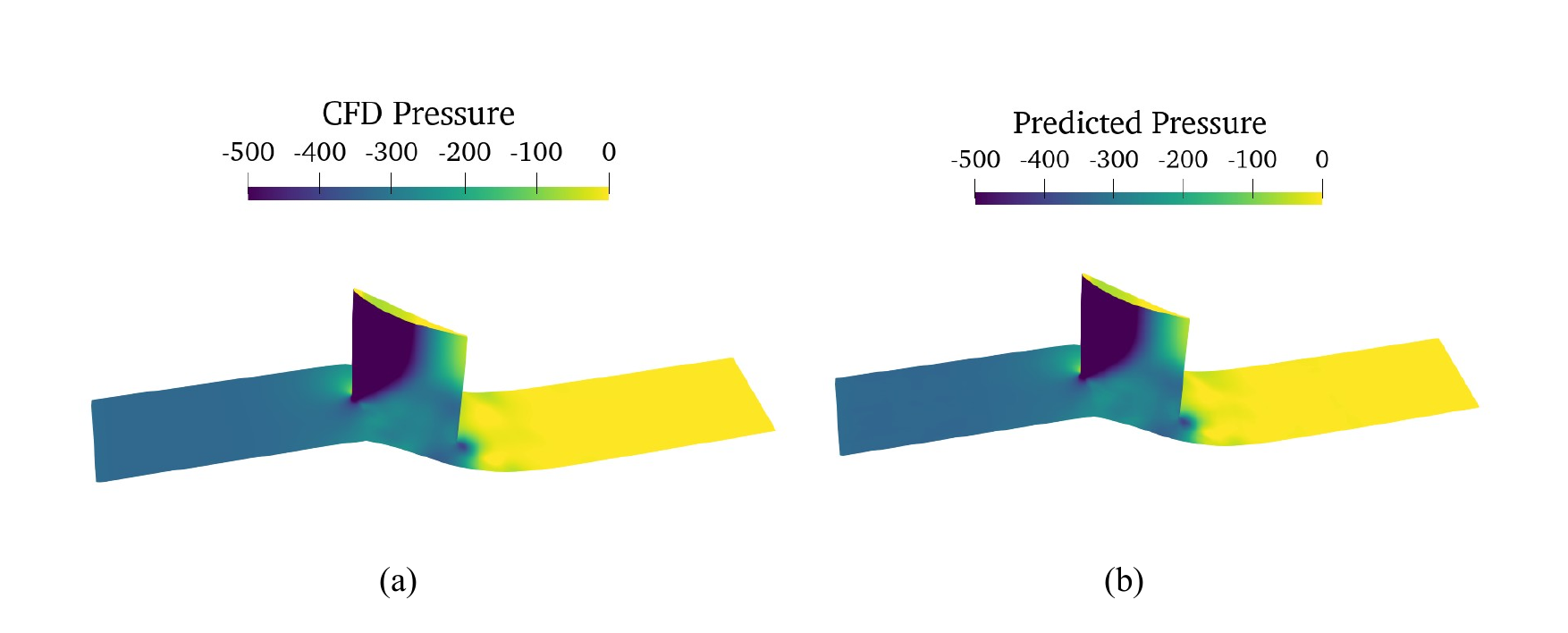}
\end{center}

\emph{Figure 4. Comparison of blade-surface pressure distributions for
the} \(5^{\circ}\)\emph{condition between the CFD reference solution and
the proposed framework}

Overall, the predicted velocity magnitude, pressure, and effective
viscosity distributions show good agreement with the CFD results. The
proposed framework successfully reconstructs the major aerodynamic
characteristics of the compressor cascade flow, including the flow
acceleration within the blade passage, the low pressure region near the
suction side, and the high viscosity regions associated with
boundary-layer development. In addition, the predicted flow fields
remain smooth and physically consistent without obvious nonphysical
oscillations.

To further evaluate the near-wall prediction performance, Fig. 4
compares the blade-surface pressure distributions between the CFD
reference solution and the predicted results. Good agreement can be
observed on both the pressure side and suction side of the blade
surface. The proposed framework accurately captures the pressure
variation along the blade surface, including regions with relatively
strong pressure gradients near the leading-edge and trailing-edge
regions. Although minor discrepancies can still be observed in localized
regions with complex three dimensional flow interactions, the overall
prediction accuracy remains satisfactory for the unseen condition. The
errors are mainly concentrated near regions with strong secondary-flow
structures and relatively large pressure gradients.

To quantitatively evaluate the prediction performance, the relative root
mean square error (RMSE) values of different flow variables are
summarized in Table 1. The results indicate that the proposed framework
achieves low prediction errors for all flow variables under the unseen
\(5^{\circ}\)attack-angle condition, demonstrating its capability for
accurate three-dimensional compressor cascade flow construction on CFD
point clouds.

\textbf{Table 1. Relative RMSE comparison of different flow variables
for the} \(\mathbf{5}^{\mathbf{\circ}}\) \textbf{prediction case}

{\def\LTcaptype{none} 
\begin{longtable}[]{@{}
  >{\centering\arraybackslash}p{(\linewidth - 10\tabcolsep) * \real{0.1599}}
  >{\centering\arraybackslash}p{(\linewidth - 10\tabcolsep) * \real{0.1599}}
  >{\centering\arraybackslash}p{(\linewidth - 10\tabcolsep) * \real{0.1599}}
  >{\centering\arraybackslash}p{(\linewidth - 10\tabcolsep) * \real{0.1599}}
  >{\centering\arraybackslash}p{(\linewidth - 10\tabcolsep) * \real{0.1600}}
  >{\centering\arraybackslash}p{(\linewidth - 10\tabcolsep) * \real{0.1600}}@{}}
\toprule\noalign{}
\begin{minipage}[b]{\linewidth}\centering
\textbf{RMSE\_\emph{u}}
\end{minipage} & \begin{minipage}[b]{\linewidth}\centering
\textbf{RMSE\_\emph{v}}
\end{minipage} & \begin{minipage}[b]{\linewidth}\centering
\textbf{RMSE\_\emph{w}}
\end{minipage} & \begin{minipage}[b]{\linewidth}\centering
\textbf{RMSE\_\emph{V\textsubscript{m}}}
\end{minipage} & \begin{minipage}[b]{\linewidth}\centering
\textbf{RMSE\_\emph{p}}
\end{minipage} & \begin{minipage}[b]{\linewidth}\centering
\textbf{RMSE\_}\(\mathbf{\nu}_{\mathbf{eff}}\)
\end{minipage} \\
\midrule\noalign{}
\endhead
\bottomrule\noalign{}
\endlastfoot
2.67％ & 0.83％ & 1.06％ & 0.74％ & 1.05％ & 3.39％ \\
\end{longtable}
}

\textbf{4 Conclusion}

In the present study, a physics-informed latent neural operator
framework was proposed for three-dimensional compressor cascade flow
prediction on CFD point clouds. The proposed method combines a
Transolver-based latent encoder with a coordinate-based PINNs decoder to
reconstruct continuous flow fields while incorporating physics-informed
residual constraints during training. The prediction performance of the
proposed framework was evaluated using the previously unseen
\(5^{\circ}\)attack-angle condition.

The results demonstrate that the proposed method can accurately
reconstruct velocity magnitude, pressure, and effective viscosity
distributions for complex three-dimensional cascade flows. Good
agreement was obtained between the predicted results and CFD reference
solutions for both global three-dimensional flow structures and
blade-surface flow distributions. In addition, the proposed framework
achieved low prediction errors for different flow variables under the
unseen operating condition.

Overall, the present study indicates that the proposed physics-informed
latent neural operator framework provides a promising approach for
efficient CFD surrogate modelling and aerodynamic prediction in
turbomachinery applications.

\textbf{References}

1. X. Guo, W. Li, and F. Iorio (2016). Convolutional neural networks for
steady flow approximation. \emph{Proceedings of the 22nd ACM SIGKDD
International Conference on Knowledge Discovery and Data Mining},
481--490.

2. O. Ronneberger, P. Fischer, and T. Brox (2015). U-Net: Convolutional
networks for biomedical image segmentation. \emph{Medical Image
Computing and Computer-Assisted Intervention (MICCAI)}, 234--241.

3. K. He, X. Zhang, S. Ren, and J. Sun (2016). Deep residual learning
for image recognition. \emph{Proceedings of the IEEE Conference on
Computer Vision and Pattern Recognition (CVPR)}, 770--778.

4. T. Pfaff, M. Fortunato, A. Sanchez-Gonzalez, and P. Battaglia (2021).
Learning mesh-based simulation with graph networks. \emph{International
Conference on Learning Representations (ICLR)}.

5. N. Brandstetter, D. Worrall, and M. Welling (2022). Message passing
neural PDE solvers. \emph{International Conference on Learning
Representations (ICLR)}.

6. Z. Li, N. Kovachki, K. Azizzadenesheli, B. Liu, K. Bhattacharya, A.
Stuart, and A. Anandkumar (2021). Fourier neural operator for parametric
partial differential equations. \emph{International Conference on
Learning Representations (ICLR)}.

7. J. Wu, W. Chen, H. Wang, et al. (2024). Transolver: A fast
transformer solver for PDEs on general geometries. \emph{Advances in
Neural Information Processing Systems (NeurIPS)}.

8. M. Raissi, P. Perdikaris, and G. E. Karniadakis (2019).
Physics-informed neural networks: A deep learning framework for solving
forward and inverse problems involving nonlinear partial differential
equations. \emph{Journal of Computational Physics}, 378:686--707.

9. W. Ma (2012). Experimental investigation of corner stall in a linear
compressor cascade. \emph{PhD thesis, École Centrale de Lyon.}

\end{document}